\documentstyle[12pt]{article}

\newcommand{\bd}{\begin{document}}
\newcommand{\ed}{\end{document}}
\newcommand{\bc}{\begin{center}}
\newcommand{\ec}{\end{center}}
\newcommand{\be}{\begin{eqnarray}}
\newcommand{\ee}{\end{eqnarray}}
\renewcommand{\thefootnote}{\alph{footnote}}
\newcommand{\se}{\section}
\newcommand{\sse}{\subsection}
\newcommand{\bi}{\bibitem}
\def\figcap{\section*{Figure Captions\markboth
     {FIGURECAPTIONS}{FIGURECAPTIONS}}\list
     {Figure \arabic{enumi}:\hfill}{\settowidth\labelwidth{Figure 999:}
     \leftmargin\labelwidth
     \advance\leftmargin\labelsep\usecounter{enumi}}}
\let\endfigcap\endlist \relax

\input psfig

\begin{document}
\baselineskip 0.73cm

\begin{titlepage}

 \vskip 0.5in
 \null
\begin{center}
 \vspace{.15in}
{\Large {\bf Implication of scalar-pseudoscalar mixing\\ on
$\epsilon'/\epsilon$ in SUSY models
}}\\
\vspace{1.0cm}  \par
 \vskip 2.1em
 {\large
  \begin{tabular}[t]{c}
{\bf Chuan-Hung Chen}
\\
\\
   {\sl Institute of Physics, Academia Sinica, Taipei,}
\\   {\sl  $\ $Taiwan 115, Republic of China }
\\

   \end{tabular}}
 \par \vskip 5.3em

\date{\today}
 {\Large\bf Abstract}
\end{center}

We study the effects of scalar-pseudoscalar mixing induced from
quantum loop on $\epsilon'/\epsilon$ in SUSY models with
$\tan\beta\sim m_{t}/m_{b}$. We find that even the non-universal
soft ${\cal A}^{d}$ term and Yukawa matrix, $Y^{d}$, are
hermitian, the predicted value of $|\epsilon'/\epsilon|$ can be
consistent with the measured results of NA48 and KTeV. And also
the EDMs are compatible with experimental bounds.
\end{titlepage}

Since its discovery in the neutral kaon decays in 1964
\cite{CCFT}, the origin of CP violation (CPV) still puzzles the
physicist. Although recently another time-dependent CP asymmetry
(CPA), $\sin 2\phi_{1}$, in the decay of $B\rightarrow J/\Psi
K_{s}$ is observed by BARBAR \cite{betaBABAR} and BELLE \cite
{betaBELLE}, our understanding of CPV  is not much better than
before. What we are certain at present is only that it is
necessary to exist unrotatable phase and it is believable that
such kind of phase is associated with weak interactions, usually
called weak phase. In the case of standard model (SM), the unique
source of CPV is from the Cabibbo-Kobayashi-Maskawa (CKM) matrix
\cite{CKM} induced from the three-generation quark mixings and
described by the three angles $\alpha $, $ \beta $ and $\gamma $
or $\phi _{2}$, $\phi _{1}$ and $\phi _{3}$.

Even though the SM prediction on the indirect CP violating
parameter $\epsilon $ in the kaon system can be fitted well with
current experimental data, due to the large uncertainties from
hadronic matrix elements, so far it has not been settled yet
whether the result in the SM can explain the observed value of the
direct CP violating parameter $\epsilon ^{\prime }$ measured by
NA48 \cite{NA48} and KTeV \cite {KTeV}. And also, because of the
unitarity in the CKM matrix, the predicted electric dipole moments
(EDMs) of neutron and lepton are quite small and unreachable
experimentally. Moreover, the requirement of Higgs boson with the
mass being less than $60$ GeV to solve the problem of baryogenesis
is ruled out by the LEP experiment. Therefore, it becomes
important to search the possibility of existing other CP violating
sources for explaining all CP phenomena.

One of reliable models in the extension of SM is supersymmetric
(SUSY) model.  SUSY theories not only supply an elegant mechanism
for the breaking of the electroweak symmetry and a solution to the
hierarchy problem, but also guarantee the unification of gauge
couplings at GUTs scale \cite{SUSY-GUTS}. In addition, SUSY
possesses abundant flavor structures, such as upper and down type
squark mixing matrices, and CP violating phases, which are arisen
from the trilinear and bilinear SUSY soft breaking ${\cal A}$ and
$B$ terms, the $\mu $ parameter for the scalar mixing etc..
Unfortunately, one can check easily that those phases are severely
bounded by electric dipole moments (EDMs) \cite{Garistosusy} so
that the effects on $\epsilon $ and $\epsilon ^{\prime }$ cannot
enough explain the current experimental values. In order to handle
the small CP phase problem, it has been suggested to use the
non-universal soft $A$ terms instead of universal ones
\cite{String1}. Furthermore, to evade any fine-tuning on the
specific phases which contribute to EDM of neutron, it is proposed
that SUSY soft-breaking ${\cal A}^{q}$ and Yukawa, $Y^{q}$,
matrices are hermitian \cite{ABKL,AKL}. Consequently, the CP
phases of $O(1)$ can exist naturally even considering the
contributions of EDM. And also, it implies that the CPA in hyperon
decays could reach the value of $O(10^{-4})$ \cite{Chen} proposed
by the experiment E871 at Fermilab \cite{White}. However, the
effect on $\epsilon'$, dominated by gluon-penguin with gluino as
the internal particle in the one-loop, will be suppressed due to
$(\delta^{d}_{12})_{LR} \simeq (\delta^{d}_{12})_{RL}$ ( the
definition is shown below). It is interesting to ask whether
$\epsilon'$, based on the gluino mechanism, can be satisfied in
the framework of hermitian ${\cal A}^{d}$ and $Y^{d}$ matrices by
considering other effects which are available after including the
constraints from experimental measurements.

According to the analysis in \cite{CQ-PLB}, we find that because
of the enhancement of large ${\cal A}^{q}$, $\mu$ and $\tan\beta$,
the coupling $N^{0}-\tilde{q}_{L}-\tilde{q}_{R}$, with $N^{0}$ and
$\tilde{q}$ being scalar (or pseudoscalar) and the squark of
corresponding to the $q$-quark respectively, will be enhanced.
Thus, it is easy to conjecture that the vanished $\epsilon'$
problem in hermitian case could be solved if we consider the
induced interactions $C_{fN^0} N^{0}\bar{f}_{Ri}f_{Lj}$ in which
$C_{fN^0}$ is the effective coupling and $f$ could be upper or
down type quarks and the indicies $i$ and $j$ denote the possible
flavour. From the induced effective vertices, if $C_{fN^0}$ is
complex, we see clearly that $\epsilon$ will also contribute. If
so, we will face the strict constraints from it. In our following
analysis, the induced effective coupling will be taken as real.
But, the CPV is generated by the scalar-pseudoscalar mixing.

It is known that if the scalar-pseudoscalar mixing comes from
spontaneous CPV, such as realized by radiative corrections in the
MSSM \cite{Maekawa}, the predicted pseudoscalar mass, $m_{A}$, is
far below the current experimental limit and excluded
\cite{Pomarol}. However, it is found that if CP is broken at the
tree level by the SUSY soft breaking sector, the large mixing
between CP-even and CP-odd boson can be obtained through radiative
corrections without the limit of small mass on the pseudoscalar
boson \cite{Pilaftsis}. As also been shown in \cite{Ellis-NPB},
the radiative CP effects will modify the couplings of Z-boson to
Higgs boson such that the mass of lightest Higgs boson with
$60-70$ GeV can escape from the bound of LEP2. With above
conclusions, in this paper, we will show their implications on
$\epsilon'$.

We start by writing the effective interactions between the
relevant neutral Higgs bosons or gluino and squarks as
 \be
{\cal L}_{eff}&=&{\cal L}_{N\tilde{f}_{L}\tilde{f}_{R}}+{\cal
L}_{\tilde{g}f\tilde{f}}, \nonumber \\
&=& \frac{g} {2M_{W}} \Big[ h^{0}{\bf \tilde{q}}^{\dagger}_{L}
\Big(({\bf M}^{2}_{\tilde{q}})_{LR}   \tilde{\omega}^{q}_{1} +
\mu^{*}{\bf m}_{q} \tilde{\omega}^{q}_{2} \Big){\bf
\tilde{q}}_{R}\nonumber \\ &&-i A^{0}{\bf
\tilde{q}}^{\dagger}_{L}\Big(  ({\bf M}^{2}_{\tilde{q}})_{LR}
 {Z}^{q}_{\beta}
- \mu^{*} {\bf m}_{q} \Big){\bf
\tilde{q}}_{R}\Big] \nonumber \\
&&-\sqrt{2}g_{s}\Big[ \bar{\bf q} P_R \tilde{g}^{a} T^{a} {\bf
\tilde{q}}_{L} -\bar{\bf q} P_L \tilde{g}^{a} T^{a} {\bf
\tilde{q}}_{R} \Big] +h.c.
 \label{leff}
 \ee
where $P_{L(R)}=(1\mp\gamma_5)/2$, $h^0$ stands for the lightest
scalar particle while $A^{0}$ is for pseudoscalar boson. The bold
${\bf q}$ and ${\bf \tilde{q}}$ denote three generaton quarks and
the corresponding squarks, respectively. The generators of
$SU(3)_c$ are normalized by $tr(T^a T^b)=1/2\delta^{ab}$.
$\tilde{\omega}^{u(d)}_{1}=\cos\alpha/\sin\beta$ $
(-\sin\alpha/\cos\beta)$ and
$\tilde{\omega}^{u(d)}_{2}=-\sin\alpha/\sin\beta$$
(\cos\alpha/\cos\beta)$ in which angle $\alpha$ describes the
mixing between two CP-even Higgs particles. For simplicity, we
only concentrate on the contributions from the lightest CP-even
Higgs. Due to the mass suppression, we expect that the effects of
heavier CP-even boson are smaller than those of light one. One the
other hand, the squared squark mass matrices responsible for
flavor change are described by
\begin{eqnarray}
{\cal M}^{2}_{\tilde{q}}=\left(\begin{array}{cc}
  ({\bf m}^2_{\tilde{q}})_{LL} & ({\bf m}^2_{\tilde{q}})_{LR}\\
 ({\bf m}^2_{\tilde{q}})^{\dagger}_{LR} & ({\bf
 m}^2_{\tilde{q}})_{RR}\\
 \end{array}\right), \label{mass}
\end{eqnarray}

\begin{eqnarray}
({\bf m}^2_{\tilde{q}})_{LL}&=&({\bf M}^2_{\tilde{q}})_{LL}+{\bf
m}^2_{q}-M^2_{Z}\cos2\beta C^{q}_{L}{\bf \hat{1}},
\nonumber \\
({\bf m}^2_{\tilde{q}})_{LR}&=&({\bf
M}^2_{\tilde{q}})_{LR}-\mu^{*} Z^{q}_{\beta} {\bf
m}_{q}, \nonumber \\
({\bf m}^2_{\tilde{q}})_{RR}&=&({\bf M}^2_{\tilde{q}})_{RR}+{\bf
m}^2_{q}+M^2_{Z}\cos2\beta C^{q}_{R}{\bf \hat{1}},
\end{eqnarray}
where we have adopted the so-called super-CKM basis that the
quarks have been the mass eigenstates so that ${\bf m}_{q}$ is the
diagonalized quark mass matrix. $q$ and $\tilde{q}$ stand for
quark and its superpartner. They could be upper or down type
quark. $Z^{u(d)}_{\beta}=\cot\beta(\tan\beta)$ and
 \be
C^{q}_{L}&=&T^3_{q}-Q_{q}\sin^2\theta_{W},\nonumber \\
C^{q}_{R}&=&Q_{q}\sin^2\theta_{W}
 \ee
with $T^{3}_{q}$ and $Q_{q}$ being the z-component of isospin
$SU(2)_{L}$ for the squark $\tilde{q}$ and its charge,
respectively. ${\bf \hat{1}}$ denotes the $3\times 3$ unit matrix.
The definition angle $\beta$ is followed by
$\tan\beta=v_{u}/v_{d}$ with $v_{u}$ and $v_{d}$ being the vacuum
expectation values (VEVs) of Higgs fields $\Phi^u$ and $\Phi^d$
responsible for the masses of upper type quarks and down type
quarks, respectively. $\mu$ is the mixing effects of $\Phi^u$ and
$\Phi^d$. $({\bf M}^2_{\tilde{q}})_{LL(RR)}$ stand for the soft
breaking masses for the corresponding squarks and $({\bf
M}^2_{\tilde{q}})_{LR}$ describe the trilinear soft breaking
couplings and are written as
\begin{eqnarray}
({\bf M}^2_{\tilde{q}})_{LR}=\frac{v_{q}}{\sqrt{2}}
V_{q_{L}}\tilde{{\cal A}}^{q\dagger} V^{\dagger}_{q_{R}},
\label{lrm}
\end{eqnarray}
where $V_{q_{L(R)}}$ transform the left(right)-handed quarks from
weak eigenstates to mass eigenstates and $\tilde{{\cal
A}}^{q}_{ij}=Y^{q}_{ij}{\cal A}^{q}_{ij}$ with $Y^{q}_{ij}$ being
the Yukawa matrix.

 For convenience, we adopt the so-called
mass-insertion approximation method \cite{Hall-NPB} in which the
basis for squark is chosen such that the gluino-squark-quark
vertices involving quarks are flavour diagonal (super-CKM basis)
instead of diagonalizing the squark mass matrix itself. Hence, the
squared squrak mass matrices are regarded as effective couplings.
If necessary, we can insert the proper effective couplings in the
propagator of squark.
%
\begin{figure}[htbp]
 \centerline{ \psfig{figure=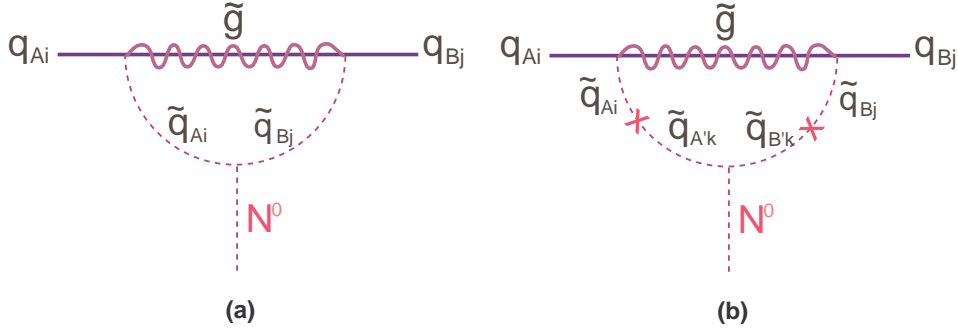,height=2.in }  } \caption{
Feynman diagrams for one-loop induced couplings of neutral Higgs
to quarks: $N^{0}$ could be scalar or pseudoscalar boson. $q$
could be upper or down type quarks, $\tilde{q}$ is the
corresponding superpartner, and the indices i,j,k stand for the
possible flavour of quarks. $A^{(')}$ and $B^{(')}$ denote the
chirality and $B^{(')}=R(L)$ while $A^{(')}=L(R)$.} \label{figha}
\end{figure}
%
According to Eqs. (\ref{leff}) and (\ref{mass}), the interactions
for $N^{0}\bar{d}_{B}s_{A}$ and $N^{0}\bar{q}_{B}q_{A}$,
illustrated in Figure \ref{figha}, can be derived as
 \be
{\cal L}&=& (4\sqrt{2}G_{F})^{1/2} \Big [(\tilde{N}^{d}_{12})_{BA}
\bar{d}_{B} s_{A} N^{0} + (\tilde{N}^{d}_{11})_{BA}
\bar{d}_{B} d_{A} N^{0} \nonumber \\
&& + (\tilde{N}^{u}_{11})_{BA} \bar{u}_{B} u_{A} N^{0} \Big]
\label{fcnc}
 \ee
where $A$ and $B$ denote the chiralities and they are always
opposite to each other,
$(\tilde{N}^{q})_{BA}=(\tilde{H}^{q})_{BA}$
($(\tilde{A}^{q})_{BA}$) if $N^{0}$ is scalar (pseudoscalar) Higgs
and their expressions are written as
\begin{eqnarray}
(\tilde{N}^{q}_{ij})_{BA} &=& \frac{\alpha_{s}}{4\pi}\sqrt{x_{q}}
C_{F} \Big[
\hat{n}^{q}_{Bij} I_{1} (x_{q}) \nonumber \\
&& + \frac{1}{3} (
\delta ^{q}_{kj})_{B'B} \hat{n}^{q}_{B'kk}   ( \delta ^{q}_{ik})_{AA'} I_{2}(x_{q}) \Big], \label{NC}\\
( \delta ^{q}_{ij})_{AA'}&=&{({\bf M}^2_{\tilde{q}ij})_{AA'} \over
m^2_{\tilde{q}}} \nonumber
\end{eqnarray}
with $x_{q}=m^2_{\tilde{g}}/m^2_{\tilde{q}}$ and $m_{\tilde{g}}$
and $m_{\tilde{q}}$ being the average masses of gluino and $q$
type squark. The first term in Eq. (\ref{NC}) is from the lowest
order contributions, illustrated in Figure \ref{figha}(a), while
the second term is generated by double mass-insertion, shown in
Figure \ref{figha}(b). The chirality $A'$ in Eq. (\ref{fcnc}) can
be $L$ or $R$, but chirality $B'$ is opposite to $A'$ so that
beside $(\delta^{q}_{ij})_{LR(RL)}$, $(\delta^{q}_{ij})_{LL(RR)}$
will also contribute.
$\hat{n}^{q}_{A}=\hat{h}^{q}_{A}(\hat{a}^{q}_{A})$ are related to
the couplings of scalar (pseudoscalar) boson to the squark
$\tilde{q}$ in which the expressions are
\begin{eqnarray*}
\hat{h}^{q}_{L}&=& \frac{({\bf
M}^{2}_{\tilde{q}})_{LR}}{m_{\tilde{q}}} \tilde{\omega}^{q}_{1} +
\frac{\mu^{*}{\bf m}_{q}}{m_{\tilde{q}}} \tilde{\omega}^{q}_{2},\\
\hat{a}^{q}_{L}&=& -i\frac{({\bf
M}^{2}_{\tilde{q}})_{LR}}{m_{\tilde{q}}} Z^{q}_{\beta} +i
\frac{\mu^{*}{\bf m}_{q}}{m_{\tilde{q}}} \label{ha}
\end{eqnarray*}
and $\hat{n}^{q}_{R}=\hat{n}^{q\dagger}_{L}$.
\begin{eqnarray*}
I_{1}(x)&=&\frac{1}{1-x}+{x \ln(x) \over (1-x)^2}, \\
I_{2}(x)&=& {2+3x-(6-x)x^2 \over 2 (1-x)^4}+ {3x\ln(x) \over
(1-x)^4}.
\end{eqnarray*}
We note that the relevant effects in the first term of Eq.
(\ref{NC}) are $\hat{h}^{d}_{L12}$, $\hat{h}^{d}_{L11}$, and
$\hat{h}^{u}_{L11}$. Via their definitions, we have
\begin{eqnarray*}
\hat{h}^{d}_{L12}&=& m_{\tilde{d}} (\delta^{d}_{12})_{LR} \tilde{\omega}^{d}_{1}, \\
\hat{h}^{d}_{L11}&=& m_{\tilde{d}} (\delta^{d}_{11})_{LR}
\tilde{\omega}^{d}_{1} + {\mu^* m_{d1} \over m_{\tilde{d}}}
\tilde{\omega}^{d}_{2}, \\
\hat{h}^{u}_{L11}&=& m_{\tilde{u}} (\delta^{u}_{11})_{LR}
\tilde{\omega}^{u}_{1} + {\mu^* m_{u1} \over m_{\tilde{u}}}
\tilde{\omega}^{u}_{2}
\end{eqnarray*}
with $m_{u1}$ and $m_{d1}$ being the masses of u-quark and
d-quark, respectively. According to the analysis in \cite{MM-PRL},
the constraint on $|Im(\delta^{d}_{12})_{LR}|$ from $\epsilon'$ is
of order of $10^{-5}$. With the assumption of the phase of $O(1)$,
we expect that the $|(\delta^{d}_{12})_{LR}|$ has the similar
order of magnitude so that its contribution is negligible.
Although it is not necessary, to simplify our analysis, we adopt
that the Yukawa matrices have the structure
$Y^{u(d)}_{33}>Y^{u(d)}_{ij}$ in which $Y^{u(d)}_{ij}$ is any
entry except $Y^{u(d)}_{33}$ and each of them is proportional to
$\sqrt{m_{i}m_{j}}/m_{3}$ with $m_{3}$ being the top or bottom
quark mass for the corresponding Yukawa matrix \cite{KKM-PRD}. As
a result, $(\delta^{q}_{11})_{LR}$ will be related to the first
two generation quark masses or be suppressed by flavour mixing
elements as defined in Eq. (\ref{lrm}). In this paper, we also
neglect their contributions. The similar situation is also applied
to $\hat{a}^{q}_{L}$. According to the above assumption, we see
that the dominant effect in second term of Eq. (\ref{NC}) is from
the component $\hat{n}^{q}_{33}$ which are expressed by
\begin{eqnarray}
\hat{h}^{q}_{L33}&=&m_{\tilde{q}}(\delta^{q}_{33})_{LR}
\tilde{\omega}^{q}_{1} + {\mu^* m_{3} \over m_{\tilde{q}}}
\tilde{\omega}^{q}_{2},\label{h33}\\
%
%
\hat{a}^{q}_{L33}&=&-im_{\tilde{q}}(\delta^{q}_{33})_{LR}
\tilde{Z}^{q}_{\beta} +i {\mu^* m_{3} \over
m_{\tilde{q}}}\label{a33}.
%
\end{eqnarray}
It is worth mentioning that due to $\hat{h}^{d}_{L33}$ and
$\hat{a}^{d}_{L33}$ associated with $1/\cos\beta$ and $\tan\beta$
respectively, if considering large $\tan\beta$ case, there is a
large enhancement.

 In terms of Eq. (\ref{fcnc}), we can immediately obtain the effective
operators for $|\Delta S|=2$ and $|\Delta S|=1$ as
 \be
{\cal L}_{|\Delta S|=2}&=&-4\sqrt{2}G_{F}\sum_{A,A'=L,R}\Big[ {
(\tilde{H}^{d}_{12})_{BA} (\tilde{H}^{d}_{12})_{B'A'}\over
m^2_{h}} \nonumber \\
&&+ {(\tilde{A}^{d}_{12})_{BA}(\tilde{A}^{d}_{12})_{B'A'} \over
m^2_{A} }  + 2{(\tilde{H}^{d}_{12})_{BA}
(\tilde{A}^{d}_{12})_{B'A'} \over m^2_{h}}\nonumber \\&& \times
\hat{\Delta}_{SP} \Big]
 \bar{d}_{B} s_{A}\, \bar{d}_{B'} s_{A'}, \label{ds=2}\\
{\cal L}_{|\Delta S|=1}&=&-4\sqrt{2}G_{F}\sum_{A,A'=L,R}\Big[
{(\tilde{H}^{d}_{12})_{BA} (\tilde{A}^{q}_{11})_{B'A'} \over
m^2_{h}} \nonumber \\
&&+{ (\tilde{A}^{d}_{12})_{BA} (\tilde{H}^{q}_{11})_{B'A'} \over
m^2_{h}} \Big]\hat{\Delta}_{SP}
 \bar{q}_{B} q_{A}\, \bar{d}_{B'} s_{A'} \label{ds=1}
 \ee
where $q=u,d$ and $\hat{\Delta}_{SP}={M^{2}_{SP}/m^2_{A}}$
describes the scalar-pseudoscalar mixing effect and its
qualitative dependence can be described by
\cite{Pilaftsis,Ellis-NPB}
\begin{eqnarray}
{\cal M}^2_{SP} \sim {m_{t}^4\over v^2} {Im(A^{t}\mu) \over
32\pi^2 M^2_{SUSY}} \left(1, \frac{|A^{t}|^2}{M^2_{SUSY}},
\frac{|\mu|^2}{\tan \beta M^2_{SUSY}},
\frac{2Re(A^{t}\mu)}{M^2_{SUSY}} \right). \label{msp}
\end{eqnarray}
In terms of the results in \cite{Pilaftsis}, one can find that one
loop radiative effects $M_{SP}$ could be of order of few hundred
GeV, that is the CP mixing factor $\hat{\Delta}_{SP}$ could be
$O(1)$. In order to study the contributions to $|\Delta S|=2$ and
$1$ decays, such as $\Delta m_{K}$, $\epsilon$ and $\epsilon'$,
the relevant hadronic matrix elements are estimated by vacuum
saturation method and written  as
  \be
{\cal K}_{1}&=&\langle K^0| \bar{d}_{R} s_{L}\, \bar{d}_{R}
s_{L}|\bar{K}^0\rangle \nonumber \\
&=& -\frac{5}{24}\Big( \frac{m_{K}}{m_s+m_d} \Big)^2 m_{K}f^2_{K},
\nonumber \\
{\cal K}_{2}&=& \langle K^0|\bar{d}_{L} s_{R}\, \bar{d}_{R} s_{L}
|\bar{K}^0\rangle \nonumber \\
&=& \Big[\frac{1}{24}+\frac{1}{4}\Big( \frac{m_{K}}{m_s+m_d}
\Big)^2\Big] m_{K}f^2_{K} \label{k12}
 \ee
for $\Delta S=2$ decays \cite{GGMS} and
 \be
{\cal P}_{1}&=&\langle\pi^{-}|\bar{d} \gamma_{5} u |0\rangle
\langle \pi^+|\bar{u}s|\bar{K}^0\rangle=-f_{\pi}B^2_{0}
\Big(1+2\frac{m^2_{\pi}}{\Lambda^2_{\chi}}\Big) \nonumber \\
{\cal P}_{2}&=&\langle \pi^{+} \pi^{-}|\bar{q} q |0\rangle \langle
0 |\bar{d} \gamma_{5} s|\bar{K}^0 \rangle
 =-f_{\pi}B^2_{0}
 \Big(1+2\frac{m^2_{K}}{\Lambda^2_{\chi}}\Big)\label{p12}
 \ee
with $B_{0}=m^2_{K}/(m_{s}+m_{d})$ and $\Lambda_{\chi}\approx 1.$
GeV for $|\Delta S|=1$ \cite{HV}. What we are concerned is only
the main effects on the FCNC decays, thus, due to $m_{h}<m_{A}$,
in the following calculations, we only concentrate on the
contributions of $m_{h}$ in Eqs. (\ref{ds=2}).

It is known that the observable for $\epsilon'/\epsilon$ is
expressed as
 \be
Re\left(\frac{\epsilon'}{\epsilon}\right)={\Omega \over
\sqrt{2}ReA_{0}|\epsilon|} (\Omega^{-1} ImA_{2}-ImA_{0})
\label{ep}
 \ee
where $A_{0}$ and $A_{2}$ denote the two-pion final states of
$K_{L}\to \pi \pi$ in isospin $I=0$ and $I=2$, respectively,
$\Omega=ReA_{2}/ReA_{0}\approx 1/22$ and $ReA_{0}\approx 2.7\times
10^{-7}$ GeV. From Eqs. (\ref{ds=1}) and (\ref{p12}), we know that
the new effects on $ImA_{2}$ and $ImA_{0}$ have similar value in
magnitude so that due to the suppressed factor $\Omega$ in isospin
$I=0$, the final state with $I=2$ is dominant. Because the CP
violating phases are arisen from soft SUSY breaking terms, in our
considering case the $(\delta^{q}_{ij})_{LL(RR)}$ are real. On the
other hand, although $(\delta^{d}_{i3})_{LR(RL)}$ ($i=1,2$) could
be complex, however, charge and color breaking (CCB) minima and
the potential unbounded from below (UFB) will give strict
constraints \cite{CD} so that their contributions are small and
negligible. In order to avoid the constraints from $\epsilon$, we
impose that $Arg(\mu)=0$, $A^{d}$ and $Y^{d}$ are hermitian
matrices so that $(\delta^{d}_{33})_{LR(RL)}$ is real, and
$(\delta^{q}_{ij})_{LL} \approx (\delta^{q}_{ij})_{RR}$ . As a
result, $\hat{h}^{d}_{A33}$ and $\hat{a}^{d}_{A33}$ are real and
purely imaginary, respectively. And then we get the identities
$(\tilde{H}^{d}_{12})_{LR}=(\tilde{H}^{d}_{12})_{RL}$ and
$(\tilde{A}^{d}_{12})_{LR}=-(\tilde{A}^{d}_{12})_{RL}$ such that
the $\epsilon$ relating effect, such as $(\tilde{H}^{d}_{12})_{RL}
(\tilde{A}^{d}_{12})_{RL} + (\tilde{H}^{d}_{12})_{LR}
(\tilde{A}^{d}_{12})_{LR}$ is vanished. Altogether, Eq.
(\ref{ds=2}) can only contribute to $\Delta m_{K}$. Moreover,
 from Eq. (\ref{msp}), we clearly see that the $M^2_{SP}$ is
related to $Im(A^{t} \mu)$, therefore, $(\delta^{u}_{33})_{LR}$ in
Eq. (\ref{ha}) should be complex. Nevertheless, in terms of Eq.
(\ref{ds=1}) and due to purely imaginary
$(\tilde{A}^{d}_{12})_{BA}$, it is obvious that only real part has
the contribution.

Combining the results of Eqs. (\ref{ds=2})$-$(\ref{ep}), we get
 \be
\Delta m_{K}&=& 2 Re\langle K^0 | {\cal H}_{|\Delta S|=2}|
\bar{K}^0 \rangle, \nonumber \\
&\sim & 16 \sqrt{2} G_{F} { (\tilde{H}^{d}_{12})^{2}_{RL} \over
m^2_{h}}
({\cal K}_{1}+{\cal K}_{2}), \label{dmk}\\
Re\left({\epsilon' \over \epsilon}\right)_{I=2}&\sim&
{2\sqrt{2}G_{F} \over 3m^{2}_{h}ReA_{0}|\epsilon|}
\Big( 1-\frac{1}{2N_c} \Big){\cal P}_{1} \nonumber \\
&& \times \Big[ (\tilde{H}^{d}_{12})_{RL}
(\tilde{A}^{u}_{11})_{RL} + (\tilde{A}^{d}_{12})_{RL}
(\tilde{H}^{u}_{11})_{RL} \Big] \hat{\Delta}_{SP} \label{dk-ep}
 \ee
where $N_{c}=3$ is the color number, for simplicity we have used
$(\hat{H}^{q}_{ij})_{RL}\approx (\hat{H}^{q}_{ij})_{LR}$ and
$(\hat{A}^{q}_{ij})_{RL} \approx -(\hat{A}^{q}_{ij})_{LR}$ for
$q=u,\, d$. Because the constraint on $(\delta^{d}_{13})_{LL}$ is
stricter than that on $(\delta^{u}_{13})_{LL}$, in  Eq.
(\ref{dk-ep}), we only show the contributions of
$(\tilde{H}^{u}_{11})_{AB}$ and $(\tilde{A}^{u}_{11})_{AB}$. To
escape the constraint from $\Delta m_{K}$ directly, we set the
${\cal A}^{b}$, $\mu$ and angle $\alpha$ satisfy with $\cot\alpha
\approx m^2_{\tilde{q}}(\delta^{d}_{33})_{RL}/\mu m_{b}$ so that
$(\tilde{H}^{d}_{12})_{RL}\approx 0$. In order to obtain the
measured value of $\epsilon'$, the values of relevant parameters
are taken as $m_{\tilde{d}}\approx m_{\tilde{u}}=m_{\tilde{q}}$,
$x_d\approx x_u=x_{q}=0.3$, $(\delta^{u}_{13})_{LR}\sim 0.1 \,
m_{\tilde{q}}/500 GeV$, $(\delta^{u}_{13})_{LL}\sim 0.3 \,
m_{\tilde{q}}/500 GeV$ \cite{BRS}, $(\delta^{d}_{23})_{LL}\sim
0.4\, (m_{\tilde{q}}/500 GeV)^2$, $(\delta^{d}_{31})_{LL}\sim 4.5
\times 10^{-2}(m_{\tilde{q}}/500 GeV)$ \cite{GGMS}, $|{\cal
A}^{t}|/m_{\tilde{q}}=|{\cal A}^{b}|/m_{\tilde{q}}\sim 2$,
$\mu/m_{\tilde{q}}\sim 1$, $(\delta^{u(d)}_{33})_{RL}\approx {\cal
A}^{t(b)} m_{b}/m^2_{\tilde{q}}$,  $m_{\tilde{q}}\approx 800$ GeV
and $\tan\beta\sim m_t/m_b$.  As mentioned before, the magnitude
of squared mass arisen from radiative corrections for the mixing
between CP-even and CP-odd boson could be order of $(100 GeV)^2$,
{\it i.e.}, $\hat{\Delta}_{SP}$ could be order of unity. As a
consequence, from Eq. (\ref{dk-ep}) and above taken values, the
predicted direct CP violating parameter for $K_{L}\to \pi \pi$ is
given by $|Re(\epsilon'/\epsilon)| \approx 1.83 \times 10^{-3}$
with $m_{h}=120$ GeV and $\hat{\Delta}_{SP}\approx 0.30$. The
result is consistent with $(15.3\pm 3.6)\times 10^{-4}$ and $(20.7
\pm 2.8)\times 10^{-4}$ given by NA48 \cite {NA48} and KTeV
\cite{KTeV}, respectively.

 Finally, we give the estimation on the electric dipole moments (EDMs) of
electron and neutron. In our present considering case, the
scalar-pseudoscalar mixing only depends on the complex
$(\delta^{u}_{33})_{LR}$. That is, even the mixing between CP-even
and CP-odd boson is small, it still can introduce CP violating
effects, such as EDMs of neutron and lepton. According to the
results of \cite{EDM-PLB}, the neutral Higgs can contribute to
EDMs of neutron and lepton via two-loop topologies. Due to
$\hat{\Delta}_{SP}$ being less than unity, the main effects should
be from pseudoscalar exchange.
 Hence, following the results of \cite{EDM-PLB}, the EDM of
fermion can be written as
\begin{eqnarray} \left(\frac{d_{f}}{e}\right)^{\gamma}
=Q_{f}{N_c \alpha_{em} \over 32\pi^3} {\tan\beta m_{f} \over
m^2_{A} }\xi_{t} Q^{2}_{t} \Big[F\Big(
\frac{m^2_{\tilde{t}_1}}{m^2_{A}} \Big) -F\Big(
\frac{m^2_{\tilde{t}_2}}{m^2_{A}} \Big) \Big]
\end{eqnarray}
where ${\tilde{t}_1}$ and ${\tilde{t}_2}$ are the mass eignestates
of stop-quark, $Q_{f}$ denotes the corresponding fermion charge
and $\xi_{t}=Z^{u}_{\beta}\, \mu\, m_{t}\,
Im(\delta^{u}_{33})_{LR}/\sin\beta \cos\beta v^2$ with
$v=\sqrt{v^2_u +v^2_d}$, the definition of function $F$ can be
found in \cite{EDM-PLB}. By using naive valence quark mode and
taking $g_{s}(\Lambda)=4\pi/\sqrt{6}$, $\alpha_{s}(M_{Z})=0.12$,
$m_{u}=7$ MeV, $m_{d}=10$ MeV, $m_{A}=400$ GeV, the predicted EDMs
of electron and neutron  with QCD renormaliztion effects,
described by
\begin{eqnarray}
\frac{d_{N}}{e} \sim \left( {g_{s}(M_{Z}) \over g_{s}(\Lambda)}
\right)^{32/23} \Big[ \frac{4}{3} \left(\frac{d_d}{e}
\right)_{\Lambda} - \frac{1}{3} \left(\frac{d_u}{e}
\right)_{\Lambda} \Big],
\end{eqnarray}
are around $1.01 \times 10^{-27}$ cm and $3.62 \times 10^{-27}$
cm, respectively. Both are satisfied the current experimental
limits given as $d_{e}/e< 4.3 \times 10^{-27}$ cm \cite{eEDM} and
$d_{N}/e< 6.3 \times 10^{-26}$ cm \cite{NEDM}.

 In summary, we have studied the effects of scalar-pseudoscalar
mixing on $\epsilon'/\epsilon$ in SUSY models with $\tan\beta\sim
m_{t}/m_{b} $. We find that if the non-universal soft ${\cal
A}^{d}$ term and Yukawa matrix, $Y^{d}$, are hermitian, and
$(\delta^{d}_{ij})_{LL}\approx (\delta^{d}_{ij})_{RR}$, the
predicted value of $\epsilon'/\epsilon$ is consistent with the
measured results of NA48 and KTeV. And also the EDMs are
compatible with experimental bounds.\\

 \noindent{\bf Acknowledgments}

I would like to thank C.Q. Geng and H.N. Li for their useful
discussions. I also thank G. Isidori for introducing me to the
constraints from CCB minima and potential UFB. This work was
supported in part by the National Science Council of the Republic
of China under Contract No. NSC-90-2112-M-001-069 and the National
Center for Theoretical Science.


\newpage
\baselineskip 0.6cm

\begin{thebibliography}{99}

\bibitem{CCFT}  J.H. Christenson, J.W. Cronin, V.L. Fitch and R. Turly,
Phys. Rev. Lett. {\bf 13}, 138 (1964).

\bibitem{betaBABAR}  BABAR Collaboration, B. Aubert {\it et al.},
Phys. Rev. Lett. {\bf 87}, 091801 (2001).

\bibitem{betaBELLE}  BELLE Collaboration, A. Abashian {\it et al.}, Phys.
Rev. Lett. {\bf 86}, 2509 (2001).


\bibitem{CKM} N. Cabibbo, Phys. Rev. Lett. {\bf 10}, 531 (1963);
 M. Kobayashi and T. Maskawa, Prog. Theor. Phys. {\bf 49}, 652 (1973)

\bibitem{KTeV}  KTeV Collaboration, A. Alavi-Harati {\it et al.}, Phys.
Rev. Lett. {\bf 83}, 22 (1999).

\bibitem{NA48} V. Fanti {\it et al.}, Phys. Lett. B{\bf 465}, 335
(1999); A. Lai {\it et al.}, Eur. Phys. J. C {\bf 22}, 231 (2001).


\bibitem{SUSY-GUTS} J. Ellis {\it et al.}, Phys. Lett. B{\bf
260}, 131 (1991); P. Langacker and M. Luo, Phys. Rev. D{\bf 44},
817 (1991).

\bibitem{Garistosusy}  R. Garisto and J.D. Wells, Phys. Rev. D{\bf 55}, 1611
(1997); R. Garisto, $ibid$, D{\bf 49} 4820 (1994) and the
references therein.

\bibitem{String1}  S.A. Abel and J.M. Fr\'{e}re, Phys. Rev. D{\bf 55}, 1623
(1997).


\bibitem{ABKL}  S. Abel, D. Bailin, S. Khalil and O. Lebedev, Phys. Lett. B{\bf 504}, 241 (2001).

\bibitem{AKL} S. Abel, S. Khalil and O.Lebedev
, hep-ph/0112260.

\bibitem{Chen} C.H. Chen, Phys. Lett. B{\bf 521}, 315 (2001).

\bibitem{White}  C. White {\it et al.}, Nucl. Phys. Proc. Suppl. B{\bf 71},
451 (1999).

\bibitem{CQ-PLB} C.H. Chen and  C.Q. Geng, Phys. Lett. B{\bf 511}, 77 (2001).

\bibitem{Maekawa} N. Maekawa, Phys. Lett. B{\bf 282}, 387 (1992).

\bibitem{Pomarol} A. Pomarol, Phys. Lett. B{\bf 287}, 331 (1992).

\bibitem{Pilaftsis} A. Pilaftsis, Phys. Lett. B{\bf 435}, 88 (1998);
Phys. Rev. D{\bf 58}, 096010 (1998); A. Pilaftsis and C.E.M.
Wagner, Nucl. Phys. B{\bf 553}, 3 (1999).

\bibitem{Ellis-NPB} M. Carena {\it et al.}, Nucl. Phys. B{\bf
586}, 92 (2000); Phys. Lett. B{\bf 495}, 155 (2000).

\bibitem{Hall-NPB} L.J. Hall, V.A. Kostelecky, and S. Rabi, Nucl.
Phys. B{\bf 267}, 415 (1986).

\bibitem{MM-PRL} A. Masiero and H. Murayama, Phys. Rev. Lett. {\bf 83}, 907
(1999).

\bibitem{KKM-PRD} S. Khalil, T. Kobayashi, and A. Masiero , Phys. Rev. D{\bf 60},
075003 (1999).

\bibitem{GGMS} F. Gabbiani {\it et al.}, Nucl. Phys. B{\bf
477}, 321 (1996).

\bibitem{HV} X.G. He and G. Valencia, Phys. Rev. D{\bf }, (1995).

\bibitem{CD} J.A. Casas and S. Dimopoulos, Phys. Lett. B{\bf 387}, 107
(1996).

\bibitem{BRS} A.J. Buras, A. Romanino and L. Silvestrini, Nucl.
Phys. B{\bf 520}, 3 (1998).


\bibitem{EDM-PLB} A. Pilaftsis, Phys. Lett. B{\bf 471},
174 (2000); D. Chang {\it et al.}, Phys. Lett. B{\bf 478}, 239
(2000).

\bibitem{eEDM} E.D. Commins, Phys. Rev. A{\bf 50}, 2960 (1994).

\bibitem{NEDM} P. G. Harris {\it et al.}, Phys. Rev. Lett. {\bf
82}, 904 (1999).

\end{thebibliography}
\end{document}